# Considerations for the design of a heterojunction bipolar transistor solar cell

Elisa Antolín, *Member, IEEE*, Marius H. Zehender,
Pablo García-Linares, Simon A. Svatek, Antonio Martí

*Abstract*— **Independent current extraction in multi-junction solar cells has gained attention in recent years because it can deliver higher annual energy yield and can work for more semiconductor material combinations than the more established series-connected multi-junction technology. The heterojunction bipolar transistor solar cell concept (HBTSC) was recently proposed as a simple, compact and cost-effective multi-terminal device structure that allows independent current extraction. It consists of only three main layers: emitter, base and collector. In this work we use a drift-diffusion model to analyze important aspects in the design of an HBTSC structure based on typical III-V semiconductor materials. We find that carrier injection from the emitter into the collector (transistor effect) degrades the open-circuit voltage of the top sub-cell, but this risk can be eliminated by optimizing the base design. We find requirements for the base layer which are, in principle, achievable in the context of current III-V semiconductor technology.**

## I. INTRODUCTION

MULTI-junction solar cells combine several sub-cells made of different semiconductor materials, each one absorbing a portion of the solar spectrum. This absorption selectivity leads to a high photovoltaic efficiency because the sub-cells converting high energy photons can deliver their photocurrent at correspondingly high voltages. This is the main reason why, multi-junction solar cells and, in particular, those made of epitaxial III-V materials, hold the absolute efficiency records to-date among all existing photovoltaic technologies [1].

In order to facilitate the integration of multi-junction devices into modules, most of them have been designed to operate under the condition of current-matching, that is, the device has only two terminals and the sub-cells are forced to produce the same photocurrent because they are interconnected in series. Although this configuration has been very successful, the current-matching constraint introduces some problems. First, current-matched structures are very sensitive to the choice of band-gap energies. Efficiency limit calculations show that adopting an independent current extraction design increases enormously the range of band-gap energies that are suitable for each different sub-cell in a multi-junction stack [2], [3]. This is a crucial aspect when there is a need to combine specific semiconductor materials for technological or economic reasons, as it is recently the case of the perovskite/silicon tandem [4], [5]. It has to be noted that, even if the chosen materials have optimum band-gap energies at 25 °C, those energies will change under real operation conditions due to temperature variations. Also, the condition of current-matching makes multi-junction solar cells sensitive to spectral changes caused by the movement of the sun in the sky and by atmospheric phenomena [6]. This results in a significant reduction of the annual energy yield for three or more junctions with respect to the case where the same junctions were independently connected [7], [8].

Those limitations in current–matched multi-junction cells have motivated a growing interest in device architectures with three or more terminals [9]–[15], in spite of the extra operational effort that they may introduce [16]–[18]. In this context, it is desirable not to introduce excessive complexity in the solar cell structure. The heterojunction bipolar transistor solar cell (HBTSC) proposed in Ref. [9] is a three-terminal device with the theoretical efficiency limit of an independently connected double-junction solar cell and with an extremely simple structure, compatible with monolithic, thin-film and low-cost technologies [19]. At this respect, it is remarkable that the HBTSC does not require the introduction of tunnel junctions or isolating layers. Fig. 1(a) shows the basic layer structure of an HBTSC. It resembles a heterojunction bipolar transistor and, as in the case of the transistor, it can be either *npn* or *pnp*. The top junction is formed between two layers called emitter (E) and base (B) in analogy to the bipolar transistor. They are made of semiconductors of high band gap energy (not necessarily the same one). The bottom junction is formed between the base (B) and collector (C) layers with the latter made of a low-bandgap semiconductor.

Manuscript received June , 2019. This work has been partially funded by the European Union's Horizon2020 Research and Innovation Programme under Grant Agreement 787289 and by the Spanish Ministry of Economy under the project INVENTA-PV (TEC2015-64189-C3-1-R). E.A. acknowledges a Ramón y Cajal Fellowship (RYC-2015-18539) funded by the Spanish Science, Innovation and Universities Ministry. Funding from Universidad Politécnica de Madrid is acknowledged by E.A. & P.G.L. (Young Researcher Grant) and M.H.Z. (Predoctoral Grant).

All authors are with the Instituto de Energía Solar - Universidad Politécnica de Madrid, 28040 Madrid, Spain (corresponding author: Elisa Antolín, elisa.antolin@upm.es).

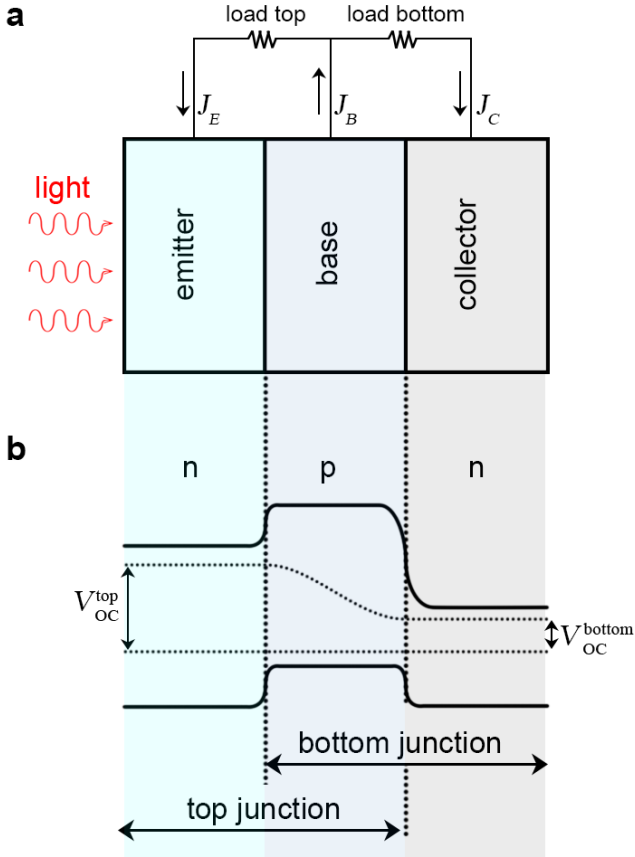

Fig. 1: (a) Structure of an *npn* HBTSC showing the three-terminals and the external circuits for independent current extraction. (b) Band diagram of an *npn* HBTSC in open circuit. Dashed lines represent the electron and hole quasi-Fermi levels. Note that the bending of the minority carrier quasi-Fermi level in the base layer enables the top junction to have a higher $V_{OC}$ than the bottom junction. The *pnp* structure is conceptually identical, exchanging the roles of electrons and holes.

Compared to conventional multi-junction technologies, the HBTSC has the advantage of a very simple structure. The current efficiency record for independently-connected double-junction devices corresponds to a four-terminal solar cell that comprises an epitaxial III-V (five layers) top cell and a complete silicon bottom solar cell, the two of them attached to either side of a glass slide [12]. On the other hand, monolithic, current-matched double-junction solar cells made of III-V semiconductors have at least 13 epitaxial layers and require a tunnel junction. An HBTSC also made of III-V materials is monolithic and contains a total of six layers, including all passivating and contact layers.

To exploit the potential of the HBTSC simple structure and make the step from theory to practical implementation it is necessary to find cost-effective and efficient device designs. In the band diagram of Fig. 1(b) we see that, conceptually, the top junction open-circuit voltage ($V_{OC}$) can be larger than the bottom junction $V_{OC}$, as expected from a double-junction solar cell, even though the two junctions share the base layer. This results from a strong bending of the minority carrier quasi-Fermi level in the base [9], [20]. The question arises: is it possible to produce this quasi-Fermi level bending without power losses and achieve a high voltage on the top cell using a base layer of realistic material parameters and reduced thickness, compatible with monolithic or low-cost technologies? In this work we apply a simple drift-diffusion model to answer this question. We use the example of an HBTSC made of III-V semiconductor materials because their properties are well characterized, but the main conclusions of the model are applicable to other material systems. We have also recently produced proof-of-concept HBTSC prototypes based on this type of materials which corroborate the main features of the HBTSC (see Ref. [18] for an AlGaAs/GaAs prototype and Ref. [21] for a GaInP/GaAs prototype).

## II. HBTSC DRIFT-DIFFUSION MODEL

In order to illustrate the role of the base layer on the performance of the HBTSC we use a drift-diffusion model. For simplification, we do not introduce illumination and study only the dark characteristics (with the exception of Fig. 6). Also, we fix all electronic parameters of the emitter and collector. Throughout the paper, only the base parameters will vary. Moreover, the thickness of emitter and collector are set to infinity. This way, their contribution to the dark current can be calculated with the long diode model (the recombination depends only on the diffusion length, and not on the actual thickness or the surface recombination velocity at the front or rear surfaces).

As long as the device is in the dark, the equations of the drift-diffusion model are the same as for an ideal bipolar transistor (BJT). It is known that the BJT performance profits from strong carrier injection from emitter to collector because this leads to a high transistor gain. For that reason, the design always considers a very small base thickness ($W_B$) and the equations are simplified. However, we can anticipate that the typical short-base approximation should not be used here because the detailed balance model of the HBTSC shows that carrier injection has to be avoided to reach the efficiency limit of a double-junction solar cell [9]. Therefore, we will use the non-approximated set of equations for a long base, which can be found in detailed BJT books such as [22]. They are compiled in Table III in the annex.

TABLE I
EMITTER AND COLLECTOR MATERIAL PARAMETERS

| | Emitter | Collector |
|---|---|---|
| Representetive of | Ga$_{0.5}$In$_{0.5}$P Al$_{0.33}$Ga$_{0.67}$As | GaAs |
| $E_G$ (eV) | 1.84 | 1.42 |
| $n_i$ (cm$^{-3}$) | 1064 | $2.1 \times 10^6$ |
| $N_D$ (cm$^{-3}$) | $2 \times 10^{17}$ | $2 \times 10^{17}$ |
| $L_h$ (μm) | 0.2 | 2 |
| $\mu_h$ (cm²/V·s) | 50 | 200 |

TABLE II
CHOICES OF MATERIAL PARAMETERS FOR THE BASE

| | BM 1 | BM2 | BM3 |
|---|---|---|---|
| Representative of | like emitter | Al-poor AlGaInP $Al_{0.5}Ga_{0.5}As$ | Al-rich AlGaInP AlAs |
| $E_G$ (eV) | 1.84 | 2.00 | 2.17 |
| $n_i$ (cm$^{-3}$) | 1064 | 274 | 12.4 |

We have chosen the material parameters so that they are approximately representative of a state-of-the-art lattice-matched III-V-semiconductor photovoltaic device. Table I summarizes the main parameters for the emitter and collector. The emitter has a band-gap energy ($E_G$) of 1.84 eV and an intrinsic carrier concentration ($n_i$) of 1064 cm$^{-3}$ at room temperature, which is representative of $Ga_{0.5}In_{0.5}P$ or $Al_{0.33}Ga_{0.67}As$ (this relationship between $E_G$ and $n_i$ is exact for AlGaAs and approximate for GaInP). The collector has the parameters of GaAs, $E_G = 1.42$ eV and $n_i = 2.1 \cdot 10^6$ cm$^{-3}$. Doping concentrations ($N_D$), minority carrier mobilities ($\mu_e$) and minority carrier diffusion lengths ($L_e$) are also given in Table I. They have been chosen so that two single-junction solar cells made of either material and stacked on top of each other would produce approximately 1.39 V and 1.02 V $V_{OC}$ under one-sun illumination.

For the base, we consider three possible materials. We have labelled them base material 1 to 3 (BM1 to BM3), as summarized in Table II. BM1 is similar to the emitter material, BM2 is representative of a higher band-gap material, such as $Al_{0.5}Ga_{0.5}As$ or AlGaInP with low Al content, and BM3 is representative of a very high band-gap material, such as AlAs or Al-rich AlGaInP. We set the base minority carrier mobility ($\mu_B$) to 100 cm$^2$/V·s and vary the doping level ($N_A$) and minority carrier diffusion length ($L_B$).

### III.  RESULTS – DESIGN OF THE BASE LAYER

Fig. 2 shows the dark current density calculated for the top junction ($J_0^{top}$) when this junction is biased at 1.3 V and the bottom junction is biased at 1.0 V (close to maximum power point operational voltages). The base thickness has been set to 800 nm and different minority carrier diffusion lengths are considered. The three plots show the three material choices for the base, from lower to higher band-gap energy: BM1 (a), BM2 (b) and BM3 (c). As for any $pn$ junction, an increase in the dark current is detrimental because it will lead to a decrease in the output power delivered by the top junction when it is under illumination (for a fixed photogenerated current density).

The $J_0^{top}$ value results from the sum of an emitter contribution (holes injected from the base, Eq. (5)) and a base contribution (electrons injected from the emitter, Eq. (4)). Therefore, the minimum possible $J_0^{top}$ is achieved when the base contribution is negligible, and it is determined by the emitter parameters. The increase in $J_0^{top}$ over that baseline that we observe in some curves can be explained by the injection of majority carriers from the emitter into the base. Fig. 2 shows that carrier injection is reduced with high band-gap energy and/or doping level in the base. Increasing the minority carrier diffusion length in the base also decreases carrier injection. It has to be noted that the plots have a logarithmic scale – any difference in $J_0^{top}$ that is noticeable to the eye in these plots will already have an impact on the one-sun $V_{OC}$ of the junction. In the case of our example with $W_B = 800$ nm we see that, if we choose material BM1, carrier injection can be avoided for very high doping levels ($\geq 10^{18}$ cm$^{-3}$) and high $L_B$ values ($\geq 300$ nm). This combination is difficult to find in practice. With material BM2 more feasible combinations work, such as ($N_A \geq 10^{17}$ cm$^{-3}$, $L_B \geq 300$ nm) or ($N_A \geq 10^{18}$ cm$^{-3}$, $L_B \geq 50$ nm). For the high band-gap material BM3, we see that carrier injection is avoided independently of other parameters.

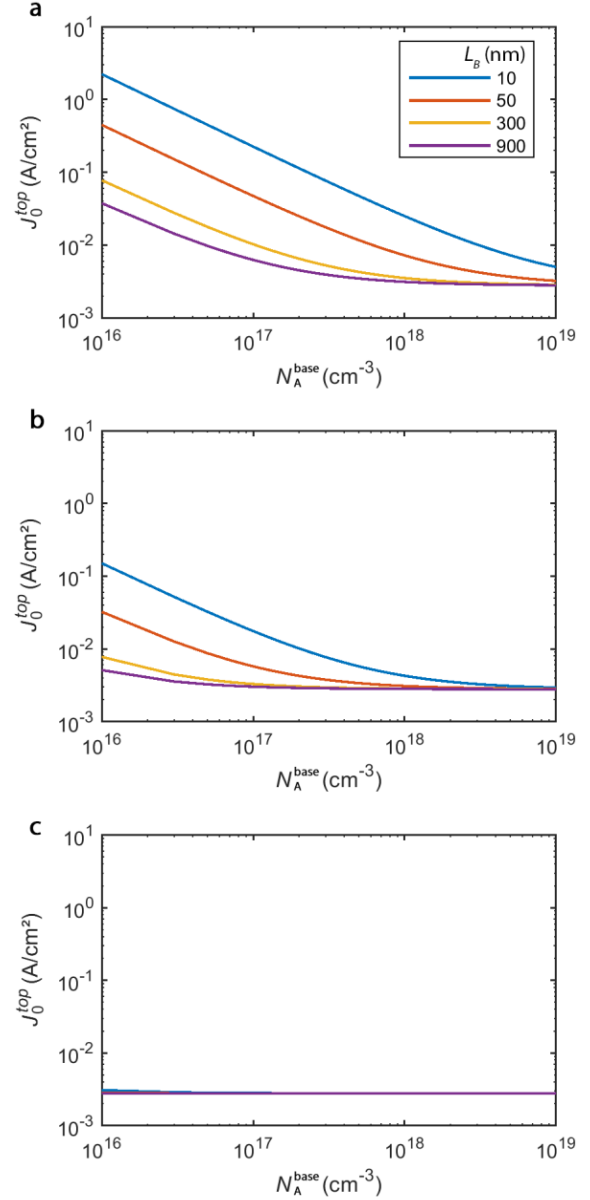

Fig. 2: Calculated dark current in the top junction at ($V_{TOP} = 1.3$ V, $V_{BOT} = 1.0$ V) vs. base doping concentration for the device described in the text ($W_B = 800$ nm) with different base materials: (a) BM1, (b) BM2 and (c) BM3. Different curves correspond to different $L_B$ values.

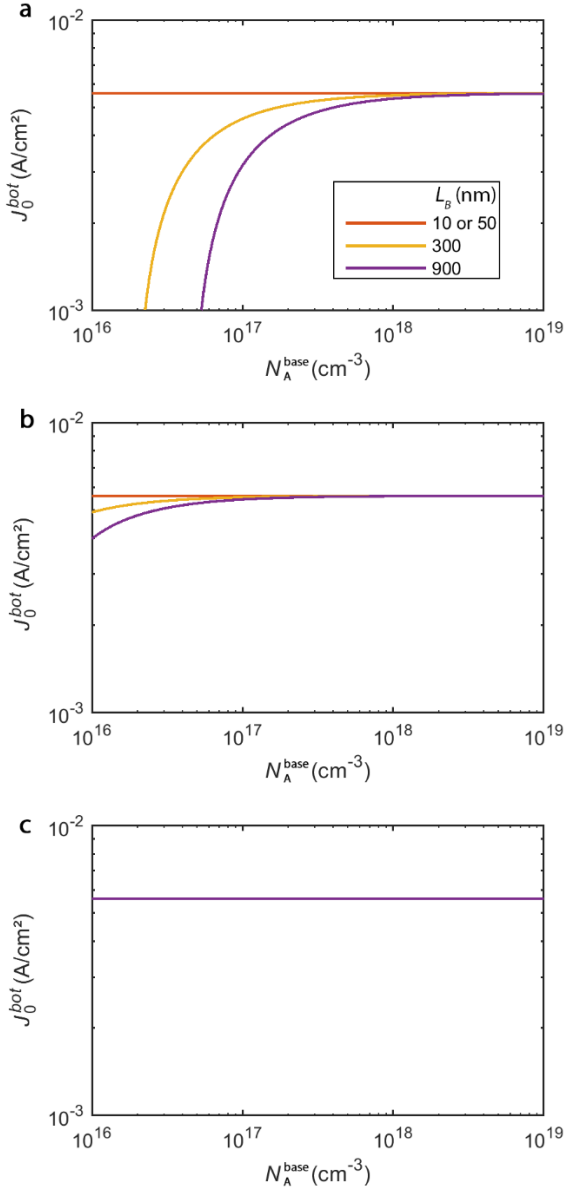

Fig. 3: Calculated dark current in the bottom junction for the same device ($W_B = 800$ nm) and at the same working point as Fig. 2, vs. base doping concentration, with different base materials: (a) BM1, (b) BM2 and (c) BM3.

Let us now discuss the dark current density in the bottom junction ($J_0^{bot}$). Fig. 3 shows $J_0^{bot}$ calculated for the same cases and the same working point as in Fig. 2. Interestingly, the combinations of band-gap energy and doping level that made $J_0^{top}$ increase, make $J_0^{bot}$ decrease. This can be understood with the aid of the sketch in Fig. 4. Since the two junctions oppose each other, they have opposite signs for the photogenerated current density ($J_L$) and the dark current density. If the dark $J_0^{top}$ injected from the emitter reaches the collector, it is seen as an apparent bottom photogenerated current density ($J_L^{bot}$) contribution. In fact, if we choose the right parameters and plot the complete dark $J_0^{bot}$ ($V_{BOT}$) curves for sufficiently high top junction voltage ($V_{top}$) values, they appear as if the junction was illuminated (see Fig. 5). Therefore, we conclude that the injection of carriers from the emitter into the collector (transistor effect) is beneficial for the photovoltaic performance of the bottom junction, although the "apparent photocurrent" added is very little compared to standard one-sun photocurrents. Note that, in principle, the bottom junction could also inject carriers into the top junction and improve its performance. However, this will not happen under normal operation conditions because $V_{top} > V_{bot}$.

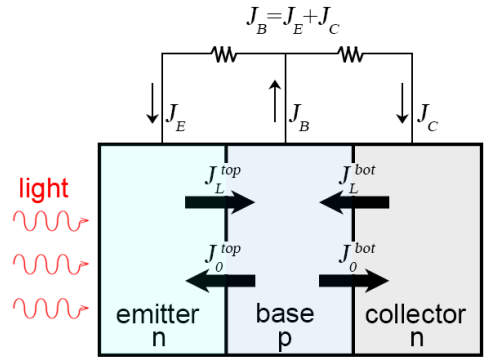

Fig. 4: Illustration of the direction of photogenerated currents and dark currents in an *npn* HBTSC.

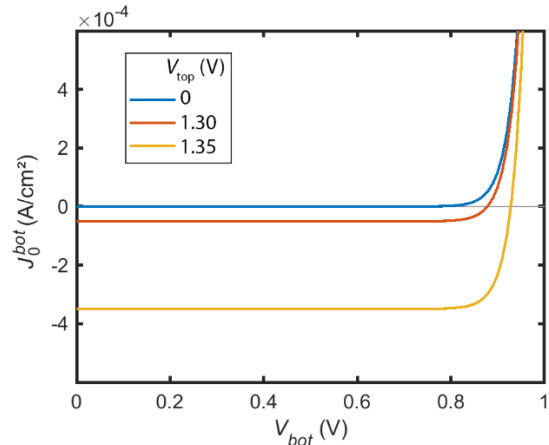

Fig. 5: Calculated bottom junction dark current for an HBTSC with base material BM1, $W_B = 800$ nm, $N_A = 7 \times 10^{17}$ cm$^{-3}$ and $L_B = 900$ nm. They look like illumination curves due to the transistor effect.

Another important aspect in the interpretation of Figs. 2 and 3 is the impact of $L_B$. The detrimental increase in $J_0^{top}$ is larger for low $L_B$ values, but the beneficial effect on $J_0^{bot}$ is stronger for high $L_B$ values. This is because, if $L_B$ is low, carrier injection from the emitter into the base is strong, but those carriers recombine in the base and do not reach the collector. Only if $L_B$ is large enough, they contribute to the performance of the bottom junction. Therefore, we can conclude that, in the cases where the band-gap energy and doping of the base allow it, small $L_B$'s degrade strongly the performance of the top junction without improving the bottom junction. Large $L_B$'s lead to a lesser degradation of the top junction and to some improvement of the bottom junction. However, when illumination currents are added to our model, the net power balance for the cell as a whole is negative because the top cell operates at a higher voltage than the bottom cell. Therefore, the loss in the top junction always exceeds the gain in the bottom junction if the two cells are biased at their respective maximum-power point voltages ($V_{MPP}$). This is consistent with the results of the detailed balance model presented in [9] which advanced that the transistor effect has to be avoided by decreasing the emitter injection efficiency in order to maximize the photoconversion efficiency. At this respect it is worth noting that when we choose the base material BM3, no carrier injection is observed for any doping level or $L_B$ value even for base thicknesses as low as 50 nm. It has to be noted that the problem of power loss due to carrier injection through the base can appear in any kind of transistor-like solar cell, not only in the heterojunction transistor like solar cell presented here. Refs. [23] and [24] present thorough discussions on the nature and the impact of minority carrier diffusion currents in the base of transistor-like silicon interdigitated back contact (IBC) solar cells designed to be used as bottom cell in current-mismatched tandems.

## IV.  IMPACT OF THE BASE DESIGN ON THE $V_{OC}$

Finally, we present in Fig. 6 an example of illumination $J$-$V$ curves for an incorrect HBTSC design in which the transistor effect has not been correctly eliminated. The following parameters have been considered: base material BM1, $W_B = 50$ nm, $N_A = 1 \times 10^{17}$ cm$^{-3}$ and $L_B = 900$ nm. This is an illustrative example where we have directly applied the superposition principle directly, subtracting from both cells a photogenerated current $J_L^{top} = J_L^{bot} = 15$ mA/cm². This approximation is acceptable if we assume that emitter and collector are able to absorb virtually all photons in their respective spectral ranges.

Fig. 6 (a) shows the illumination $J$-$V$ curve of the top junction. We have considered the cases where the bottom junction is short-circuited (red curve) and biased at its $V_{MPP}$ (orange curve). Note that in cases where there is cross-talk between the junctions, such as here, the MPP has to be calculated by maximizing the total power produced by both junctions. As a reference, we have also plotted the case of an isolated homojunction with emitter and base having the same parameters as before and an infinite length (long diode).

In Fig. 6 (b) similar curves are plotted for the bottom junction. To be consistent in the comparison with reference devices, in this case the homojunction considered would be a GaAs homojunction (blue curve). This way, both blue curves together illustrate the case of a conventional four-terminal double-junction solar cell.

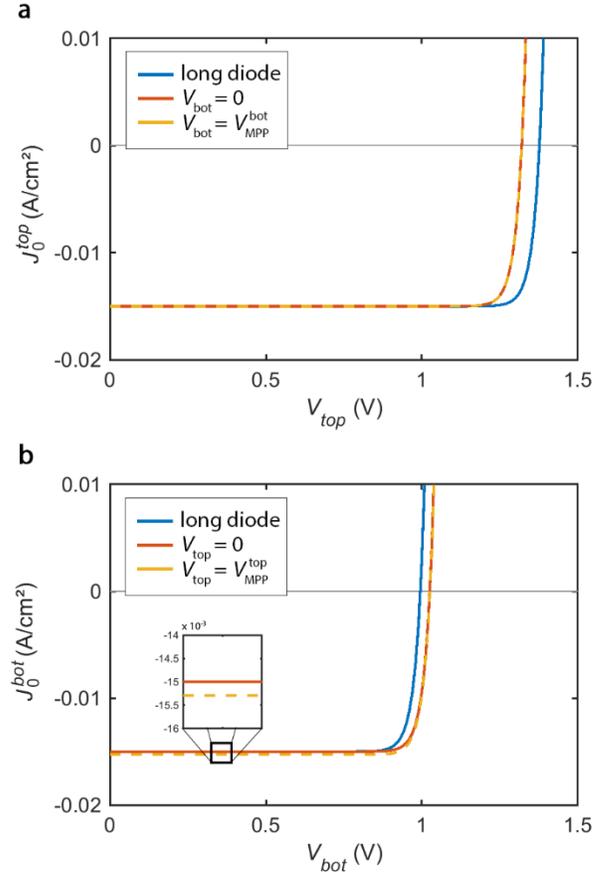

Fig. 6: Illumination $J$-$V$ curves calculated for an incorrectly designed HBTSC where carrier injection from emitter to collector is possible. (a) top junction, (b) bottom junction. The red and orange curves represent different voltages of the opposite junction. The blue curves represent, respectively, the top and bottom homojunctions in a conventional independently-connected double-junction solar cell.

We observe that the photocurrent of the bottom cell increases slightly when $V_{top}$ is raised [see inset in Fig. 6 (b)], which can be explained with the same arguments applied to Fig. 5. Changes in the top junction plot are not appreciable when $V_{bot}$ is varied. What results very noticeable in this example of non-optimal HBTSC design is that the $V_{OC}$ in the top cell is worse than in the reference homojunction. Having a thin base layer, the top junction of the HBTSC behaves like a short diode which has not been passivated, which compares badly to a long diode with the same material parameters. In the case of the bottom junction, we see that the non-optimal HBTSC design has been beneficial, because the base component to the dark current is smaller than in a GaAs homojunction. However, the total balance when both junctions are considered is negative for the non-optimal HBTSC. Having allowed carrier injection comes at the cost of a decrease in efficiency with respect to the conventional double-junction design. Nevertheless, it has to be noted that the total efficiency loss is marginal, from 31.92 % in the long diode case to 31.80 % in the non-optimized HBTSC. Such a small decrease is striking if we take into account that the base thickness in this non-optimized example is extremely low (50 nm) and the doping level quite low ($1 \times 10^{17}$ cm$^{-3}$). This illustrates that the HBTSC design is very robust with respect to carrier injection, provided that the band-gap energy of the base material is equal or greater than of the emitter material.

## V. CONCLUSION

Using a simple drift-diffusion model with parameters representative of III-V materials, we have shown that the penalty that can be expected from the compact structure of the HBTSC is a voltage loss (dark current increase) in the top cell if the design is not optimized. This voltage loss is related to the injection of carriers from the emitter into the base. If the carriers reach the collector (long minority carrier diffusion length in the base), the performance of the bottom junction can be marginally improved. However, the overall balance between degradation of the top cell and improvement of the bottom cell is always negative for the ideal devices studied here when carrier injection is possible. Therefore, maximum efficiency can only be achieved when that affect is eliminated.

Further indications are given about the suitable base parameter space to design an efficient III-V semiconductor HBTSC. In general, it is found that increasing the band-gap energy, the doping concentration and/or the minority carrier diffusion length in the base reduces carrier injection and improves the overall performance. Some combinations of those parameters are feasible from the point of view of III-V material technology. In particular, it is found that, if the material in the base has a sufficiently high band-gap energy (above 2.1 eV), carrier injection is eliminated independently of all other base parameters, and, specifically, independently of having a thin base layer. In fact, our calculations show that with that band-gap energy carrier injection is negligible even for a base as thin as 50 nm. This implies that the HBTSC concept is compatible with compact, thin-film and cost-effective technologies.

## VI. REFERENCES

[1] F. Dimroth *et al.*, "Four-Junction Wafer-Bonded Concentrator Solar Cells," *IEEE J. Photovolt.*, vol. 6, no. 1, pp. 343–349, Jan. 2016.

[2] Z. (Jason) Yu, M. Leilaeioun, and Z. Holman, "Selecting tandem partners for silicon solar cells," *Nat. Energy*, vol. 1, p. 16137, Sep. 2016.

[3] M. W. Wanlass, K. A. Emery, T. A. Gessert, G. S. Horner, C. R. Osterwald, and T. J. Coutts, "Practical considerations in tandem cell modeling," *Sol. Cells*, vol. 27, no. 1, pp. 191–204, Oct. 1989.

[4] M. T. Hörantner and H. J. Snaith, "Predicting and optimising the energy yield of perovskite-on-silicon tandem solar cells under real world conditions," *Energy Environ. Sci.*, vol. 10, no. 9, pp. 1983–1993, Sep. 2017.

[5] J. Werner, B. Niesen, and C. Ballif, "Perovskite/Silicon Tandem Solar Cells: Marriage of Convenience or True Love Story? – An Overview," *Adv. Mater. Interfaces*, vol. 5, no. 1, 1700731, 2018.

[6] A. S. Brown and M. A. Green, "Limiting efficiency for current-constrained two-terminal tandem cell stacks,"

[7] J. Villa and A. Martí, "Impact of the Spectrum in the Annual Energy Production of Multijunction Solar Cells," *IEEE J. Photovolt.*, vol. 7, no. 5, pp. 1479–1484, Sep. 2017.

[8] H. Schulte-Huxel, T. J. Silverman, M. G. Deceglie, D. J. Friedman, and A. C. Tamboli, "Energy Yield Analysis of Multiterminal Si-Based Tandem Solar Cells," *IEEE J. Photovolt.*, vol. 8, no. 5, pp. 1376–1383, Sep. 2018.

[9] A. Martí and A. Luque, "Three-terminal heterojunction bipolar transistor solar cell for high-efficiency photovoltaic conversion," *Nat. Commun.*, vol. 6, p. 6902, Apr. 2015.

[10] M. A. Steiner *et al.*, "A monolithic three-terminal GaInAsP/GaInAs tandem solar cell," *Prog. Photovolt. Res. Appl.*, vol. 17, no. 8, pp. 587–593, 2009.

[11] T. Nagashima, K. Okumura, K. Murata, and Y. Kimura, "Three-terminal tandem solar cells with a back-contact type bottom cell," in *Conference Record of the Twenty-Eighth IEEE Photovoltaic Specialists Conference - 2000 (Cat. No.00CH37036)*, 2000, pp. 1193–1196.

[12] S. Essig *et al.*, "Raising the one-sun conversion efficiency of III–V/Si solar cells to 32.8% for two junctions and 35.9% for three junctions," *Nat. Energy*, vol. 2, no. 9, p. 17144, Sep. 2017.

[13] E. L. Warren, M. G. Deceglie, M. Rienäcker, R. Peibst, A. C. Tamboli, and P. Stradins, "Maximizing tandem solar cell power extraction using a three-terminal design," *Sustain. Energy Fuels*, vol. 2, no. 6, pp. 1141–1147, May 2018.

[14] M. Schnabel *et al.*, "Equivalent Performance in Three-Terminal and Four-Terminal Tandem Solar Cells," *IEEE J. Photovolt.*, vol. 8, no. 6, pp. 1584–1589, Nov. 2018.

[15] S. Sista, Z. Hong, M.-H. Park, Z. Xu, and Y. Yang, "High-Efficiency Polymer Tandem Solar Cells with Three-Terminal Structure," *Adv. Mater.*, vol. 22, no. 8, pp. E77–E80, 2010.

[16] J. M. Gee, "A comparison of different module configurations for multi-band-gap solar cells," *Sol. Cells*, vol. 24, no. 1, pp. 147–155, May 1988.

[17] H. Schulte-Huxel, D. J. Friedman, and A. C. Tamboli, "String-Level Modeling of Two, Three, and Four Terminal Si-Based Tandem Modules," *IEEE J. Photovolt.*, vol. 8, no. 5, pp. 1370–1375, Sep. 2018.

[18] M. Zehender *et al.*, "Module interconnection for the three-terminal heterojunction bipolar transistor solar cell," *AIP Conf. Proc.*, vol. 2012, no. 1, p. 040013, Sep. 2018.

[19] P. G. Linares, E. Antolín, and A. Martí, "Novel heterojunction bipolar transistor architectures for the practical implementation of high-efficiency three-terminal solar cells," *Sol. Energy Mater. Sol. Cells*, vol. 194, pp. 54–61, Jun. 2019.

[20] A. Martí, E. Antolín, P. García-Linares, E. López, J. Villa, and I. Ramiro, "Operation of the Three Terminal Heterojunction Bipolar Transistor Solar Cell," *Phys. Status Solidi C*, vol. 14, no. 10, p. 1700191, 2017.

*Prog Photovolt Res Appl*, vol. 10, no. 5, pp. 299–307, 2002.

[21] M. H. Zehender *et al.*, "Demonstrating the GaInP/GaAs Three-Terminal Heterojunction Bipolar Transistor Solar Cell," presented at the 46th IEEE Photovoltaics Specialists Conference, Chicago, 2019.

[22] G. W. Neudeck, *The bipolar junction transistor*, vol. III. Adison-Wesley.

[23] R. Rienäcker, E. L. Warren, T. F. Wietler, P. Stradins, A. C. Tamboli, and R. Peibst, "Three-Terminal Bipolar Junction Bottom Cell as Simple as PERC: Towards Lean Tandem Cell Processing," in *Proceedings of the 46th IEEE PVSC*, 2019.

[24] P. Stradins, M. Rienaecker, R. Peibst, A. C. Tamboli, and E. L. Warren, "A simple physical model for three-terminal tandem cell operation," in *Proceedings of the 46th IEEE PVSC*, 2019.

## VII. ANNEX

We summarize in Table III the equations used in this work. The first four are definitions and the rest constitute the theoretical model. They are the solution to the system of current drift-diffusion equations, continuity equations and the Poisson equation when they are applied to the emitter, base and collector. We have assumed that all regions are electrically long. $k$ is the Boltzmann constant and $T$ the absolute temperature. The derivation of this model can be found in texts where the BJT is explained in detail, such as [22].

TABLE III
EQUATIONS OF THE DRIFT-DIFFUSION MODEL

$$P_{OE(C)} = \frac{n_{iE(C)}^2}{N_{D,E(C)}} \ , \ n_{OB} = \frac{n_{iB}^2}{N_{DB}} \tag{1}$$

$$n_i^2 \propto exp\left(\frac{-E_G}{kT}\right) \tag{2}$$

$$D_X = \frac{kT}{e}\mu_X \tag{3}$$

$$F(V) = \exp\left(\frac{eV}{kT}\right) - 1 \tag{4}$$

$$J_{En}(V_{BE}, V_{BC}) = \frac{eD_B n_{0B}}{L_B}\left[\left(\coth\left(\frac{W_B}{L_B}\right)\cdot F(V_{BE})\right) - \left(\frac{1}{\sinh(W_B/L_B)}\cdot F(V_{BC})\right)\right] \tag{5}$$

$$J_{Ep}(V_{BE}) = \frac{eD_E P_{OE}}{L_E}F(V_{BE}) \tag{6}$$

$$J_E = J_{En} + J_{Ep} = J_0^{top} \tag{7}$$

$$J_{Cn}(V_{BE}, V_{BC}) = \frac{eD_B n_{0B}}{L_B}\left[\left(\frac{1}{\sinh(W_B/L_B)}\cdot F(V_{BE})\right) - \left(\coth\left(\frac{W_B}{L_B}\right)\cdot F(V_{BC})\right)\right] \tag{8}$$

$$J_{Cp}(V_{BC}) = -\frac{eD_C P_{OC}}{L_C}F(V_{BC}) \tag{9}$$

$$J_C = J_{Cn} + J_{Cp} = -J_0^{bot} \tag{10}$$

$$J_B = J_E + J_C \tag{11}$$